\documentclass[
  journal=jacsat,
  layout=twocolumn,
  manuscript=communication,
  biblabel=fullstop
]{achemso}

\usepackage[utf8]{inputenc}
\usepackage{braket}
\graphicspath{{fig/}}

\usepackage{amsmath,amssymb,units,txfonts}
\usepackage{amsfonts}
\usepackage{helvet,wasysym}
\usepackage[usenames,dvipsnames]{color}
\usepackage{colortbl,cuted}
\usepackage{multirow}

\definecolor{JPCCBlue}{RGB}{34,80,169}
\definecolor{NLRed}{RGB}{215,24,30}
\definecolor{abstractcolor}{RGB}{255,243,201}
\makeatletter\newenvironment{abstractbox}{%
   \begin{lrbox}{\@tempboxa}\begin{minipage}{0.988\textwidth}}{\end{minipage}\end{lrbox}%
   \colorbox{abstractcolor}{\usebox{\@tempboxa}}
}\makeatother

\renewcommand*\authorfont{\flushleft}
\renewcommand*\affilfont{\mdseries\flushleft\textrm}
\renewcommand*\titlesize{\Large}
\renewcommand*\titlefont{\flushleft\sf\bfseries}
\def\refname{\sffamily\color{JPCCBlue}\Large$\blacksquare$\normalsize\ REFERENCES}
\usepackage{titlesec}
\titleformat{\section}{\bfseries\sffamily\color{JPCCBlue}}{\thesection.~}{0pt}{}
\titleformat{\subsection}[runin]{\bfseries\sffamily\normalsize}{\indent\thesubsection.~}{0pt}{}[.]
\titlespacing{\subsection}{0pt}{0pt}{*1}
\titleformat{\subsubsection}{\bfseries\sffamily\normalsize}{\thethesubsection.~}{0pt}{}
\titlespacing{\subsubsection}{0pt}{0pt}{*0}

\newcommand{\ve}[1]{\mathbf{#1}}
\newcommand{\ee}[0]{\mathrm{e}}
\newcommand{\dee}[0]{\mathrm{d}}
\newcommand{\idee}[0]{\,\dee}
\newcommand{\Ha}[0]{\mathrm{H}}
\newcommand{\ext}[0]{\mathrm{ext}}

\newcommand{\XC}[0]{\mathrm{xc}}
\newcommand{\X}[0]{\mathrm{x}}

\newcommand{\op}[1]{\hat{#1}}

\newcommand{\diff}[2]{\frac{\dee#1}{\dee#2}}

\DeclareMathOperator{\real}{Re}
\DeclareMathOperator{\imag}{Im}

\graphicspath{{fig/}}

\title{Stark Ionization of Atoms and Molecules within Density Functional Resonance Theory}

\author{Ask Hjorth Larsen}
\email{asklarsen@gmail.com}

\author{Umberto De Giovannini}
\email{umberto.degiovannini@ehu.es}

\affiliation[UPV/EHU]{\footnotemark[2]{\ } Nano-Bio Spectroscopy Group and ETSF Scientific
  Development Centre, Departamento de Física de Materiales,
  Universidad del País Vasco, CSIC-UPV/EHU-MPC and DIPC, Avenida de
  Tolosa 72, E-20018 San Sebastián, Spain}

\author{Daniel L. Whitenack}

\affiliation[Purdue]{\newline\footnotemark[3]{\ } 
  Department of Physics, Purdue University, 525
  Northwestern Avenue, West Lafayette, IN 47907, USA}

\author{Adam Wasserman}
\email{awasser@purdue.edu}

\affiliation[Purdue]{\newline\footnotemark[5]{\ } Department of Chemistry, Purdue University, 560 Oval Drive, West Lafayette, IN 47907, USA}
\alsoaffiliation[Purdue]{\newline\footnotemark[3]{\ } 
  Department of Physics, Purdue University, 525
  Northwestern Avenue, West Lafayette, IN 47907, USA}

\author{Angel Rubio}
\email{angel.rubio@ehu.es}
\affiliation[UPV/EHU]{\footnotemark[2]{\ } Nano-Bio Spectroscopy Group and ETSF Scientific
  Development Centre, Departamento de Física de Materiales,
  Universidad del País Vasco, CSIC-UPV/EHU-MPC and DIPC, Avenida de
  Tolosa 72, E-20018 San Sebastián, Spain}

\begin{document}

\maketitle

\begin{strip}
\vspace{-1.6cm}

\noindent{\color{JPCCBlue}{\rule{\textwidth}{0.5pt}}}
\begin{abstractbox}
\begin{tabular*}{17cm}{b{11.0cm}r}
\noindent\textbf{\color{JPCCBlue}{ABSTRACT:}}
  We show that the energetics and lifetimes of resonances of finite
  systems under an external electric field can be captured by
  Kohn--Sham density functional theory (DFT) within the formalism of
  uniform complex scaling.  Properties of resonances are calculated
  self-consistently in terms of complex densities, potentials and
  wavefunctions using adapted versions of the known algorithms from
  DFT.  We illustrate this new formalism by calculating ionization rates
  using the complex-scaled local density
  approximation and exact exchange.  We consider a variety of
  atoms (H, He, Li and Be) as well as the H$_2$ molecule.
  Extensions are briefly discussed.\newline
\noindent \textbf{\color{JPCCBlue}{SECTION:}}  Spectroscopy, Photochemistry, and Excited States\newline
\noindent \textbf{\color{JPCCBlue}{KEYWORDS:}} Resonances, Tunneling, Spectroscopy, Lasers, Excitations, Complex scaling, Open quantum systems.\newline

&\colorbox{white}{
\includegraphics[width=4.94cm]{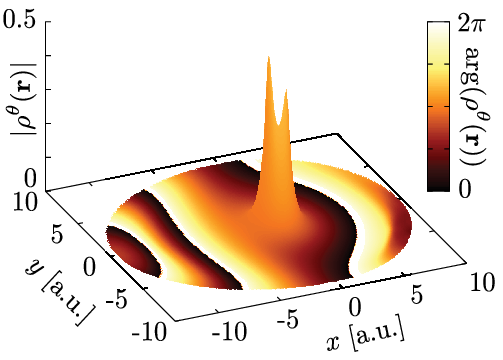}}\\
\end{tabular*}
\end{abstractbox}
\noindent{\color{JPCCBlue}{\rule{\textwidth}{0.5pt}}}

{\scriptsize This document is the unedited Author's version of a Submitted Work that was
subsequently accepted for
publication in J.Phys.Chem.Lett., copyright © American Chemical Society after
peer review. To access the final edited
and published work see \texttt{http://pubs.acs.org/doi/abs/10.1021/jz401110h}.}
\end{strip}

\def\bigfirstletter#1#2{{\noindent
    \setbox0\hbox{{\color{JPCCBlue}{\Huge \rm{#1}}}}\setbox1\hbox{#2}\setbox2\hbox{(}%
    \count0=\ht0\advance\count0 by\dp0\count1\baselineskip
    \advance\count0 by-\ht1\advance\count0 by\ht2
    \dimen1=.5ex\advance\count0 by\dimen1\divide\count0 by\count1
    \advance\count0 by1\dimen0\wd0
    \advance\dimen0 by.25em\dimen1=\ht0\advance\dimen1 by-\ht1
    \global\hangindent\dimen0\global\hangafter-\count0
    \hskip-\dimen0\setbox0\hbox to\dimen0{\raise-\dimen1\box0\hss}%
    \dp0=0in\ht0=0in\box0}#2}

\bigfirstletter
The description of metastable compounds has been elusive to first-principles
calculations due to the lack of a variational
principle. The concepts of metastability and long-lived resonances (or
tunneling processes) are closely related, and in the end we are facing the
description of the lifetime of a given open quantum system.
One approach to such calculations is the
complex-scaling method,
pioneered by Aguilar, Balslev and Combes\cite{aguilarcombes,balslevcombes}.
Within this formalism, resonances appear as the result of a
complex scaling $\ve r \rightarrow \ve r \ee^{i\theta}$ of the real-space
coordinates in the Hamiltonian.  The method has been used to calculate
resonance energies and lifetimes of negative ions of
atoms\cite{PhysRevA.20.814},
as well as resonances induced
by static electric fields\cite{QUA:QUA560100840,PhysRevLett.41.67}.

Applications have however generally been limited to small systems
or systems with reduced dimensionality
due to the computational difficulty of solving many-particle problems.
A different approach must be taken to accommodate realistic systems
with many electrons.
Recently it
has been proven that the low-lying metastable states of a given system
can be described within a density functional framework once we allow
for complex densities\cite{WM07}.
An analog of the Hohenberg--Kohn theorem then allows 
for the calculation of the
lowest-energy resonance of a system.  Based on this, the first Kohn--Sham
density functional resonance theory (DFRT)
calculations
have since been published, although limited to 1D
systems with one or two electrons\cite{doi:10.1021/jz9001778,WW11}.

A notable ongoing development, termed complex DFT (CODFT), is based on
complex absorbing potentials\cite{doi:10.1021/jz3006805,zhou:094105}.
This method relies on the definition of an absorption zone outside the
system boundary to calculate lifetimes based on how wavefunctions
extend into the absorbing region.
Another method is exterior complex scaling,
where a complex coordinate scaling is applied outside of a certain
radius\cite{Simon1979211}.

The method presented in this Letter is based on
uniform complex scaling, where all regions of space are treated
equally.  We present first-principles DFRT calculations of Stark resonance
states and lifetimes in real 3D systems within exact exchange (EXX)
and the local density
approximation (LDA).  We consider the H, He,
Li, and Be atoms and the H$_2$ molecule within strong electric fields,
extending the method beyond the model systems for which it has been
demonstrated previously\cite{doi:10.1021/jz9001778,WW11}.
The implementation is open source and part of the DFT code
Octopus\cite{Marques200360,PSSB:PSSB200642067}. Wavefunctions are
represented on real-space grids, and atoms are represented by
pseudopotentials.  Atomic units are used throughout this article.

We describe first how complex scaling is incorporated
within DFT, then present the major algorithmic steps involved.
Finally we discuss the results.

The complex-scaling transformation $\ve r \rightarrow \ve r
\ee^{i\theta}$ changes the Hamiltonian of a system into a non-Hermitian operator
$\op H^\theta$, affecting bound and unbound eigenstates differently.
The energy of any bound state is conserved, while
that of a non-normalizable, unbound state changes.
As $\theta$ is increased from 0, resonances can be
uncovered from among the continuum 
as localized eigenstates of $\op H^\theta$.
From the corresponding complex eigenvalues
$\epsilon=\epsilon_{\mathrm{res}} - i \Gamma / 2$,
the ionization rate is given by $\Gamma$; see e.g.\ the review by
Reinhardt\cite{doi:10.1146/annurev.pc.33.100182.001255}.

Standard Kohn--Sham (KS) DFT is formulated as the minimization of an energy
functional over a
set of auxiliary single-particle states.
Correspondingly we take the complex-valued Kohn--Sham energy
functional\cite{WW11}
to be
\begin{align}
  E_{\mathrm{res}} - i\frac{\Gamma}{2} &= \ee^{-i2\theta}
  \sum_n \int \psi_n^\theta(\ve r) \left(-\frac12\nabla^2\right) 
  \psi_n^\theta(\ve r) \idee \ve r\nonumber\\
  &\quad+ \ee^{-i\theta}\frac12\int\hspace{-6pt}\int
  \frac{n^\theta(\ve r) n^\theta(\ve r')}{\Vert \ve r - \ve r'\Vert}
  \idee \ve r \idee \ve r'\nonumber\\
  &\quad+ E_\XC^\theta[n^\theta]
  +\int v_\ext^\theta(\ve r) n^\theta(\ve r) \idee \ve r
  \label{eq:complexenergy},
\end{align}
where $E_{\mathrm{res}}$ is the resonance energy and $\Gamma$ the ionization rate 
inverse lifetime.  $\Gamma$ represents a total over all metastable
single-electron states in the system.  Above we have introduced
complex-scaled Kohn--Sham states $\psi_n^\theta(\ve r)$,
density $n^\theta(\ve r)$, and
operators $\op O^\theta(\ve r)$
(such as $v_\ext^\theta(\ve r)$ and the kinetic operator),
which are analytic continuations
\begin{align}
  \psi^\theta(\ve r)&=\ee^{i3\theta/2}\psi(\ve r\ee^{i\theta}),\label{eq:csstates}\\
  n^\theta(\ve r) &= \sum_n [\psi_n^\theta(\ve r)]^2 =
  \ee^{i3\theta} n(\ve r \ee^{i\theta})
  \label{eq:csdensity}\\
  \op O^\theta(\ve r)&=\op O(\ve r\ee^{i\theta}),\label{eq:csoperators}
\end{align}
of their unscaled equivalents.
We have used that bra states are not conjugated\cite{doi:10.1080/00268977800102631},
and that left and right eigenstates are
equal, since the complex-scaled Hamiltonian of a
finite system is complex-symmetric.
This is always the case when the initial,
unscaled Hamiltonian contains only real terms.
These definitions ensure that
integrals such as matrix elements
are unaffected by the scaling angle $\theta$ under appropriate
conditions\cite{doi:10.1146/annurev.pc.33.100182.001255}.
Thus \eqref{eq:complexenergy} reduces
to the ordinary KS energy functional if the system is bound.

For unbound systems the energy functional is complex and therefore
does not possess a minimum.  However the lowest resonance can still
be obtained by requiring the functional to be stationary.\cite{WM07}
Taking the derivative with respect to the wavefunctions
$\psi_n^\theta(\ve r)$ yields a set of KS equations:
\begin{align}
  \op H^\theta \psi_n^\theta(\ve r) =
  \left[-\frac12 \ee^{-i2\theta} \nabla^2 + v^\theta(\ve r)\right] 
  \psi_n^\theta(\ve r)
  = \epsilon_n \psi_n^\theta(\ve r),\label{eq:kohnsham}
\end{align}
where the effective potential $v^\theta(\ve r)$ is the sum
of the complex-scaled Hartree, XC and external potentials\cite{WW11}
$v^\theta(\ve r) = v_\Ha^\theta(\ve r) 
     + v_\XC^\theta(\ve r) + v_\ext^\theta(\ve r)$.
A self-consistency loop is then formulated from these
quantities.
For each calculation a fixed value of $\theta$ is chosen.
$\theta$ should be large enough for the resonant states to emerge,
but can otherwise be chosen to optimize the
numerics.\cite{QUA:QUA560140408,whitenack:164106}
In the DFRT calculations presented here, $\theta$ is
chosen by testing different values with different grid
spacings to find a combination with good numerical precision.
This is similar to the convergence checks performed in standard DFT,
but more important, as the numerical error must be made smaller than the
imaginary part of the energy, which can be quite close to zero.  For this
reason a fine grid is required to calculate low ionization rates.

\begin{figure}
  \noindent{\color{JPCCBlue}{\rule{\columnwidth}{1pt}}}
  \includegraphics{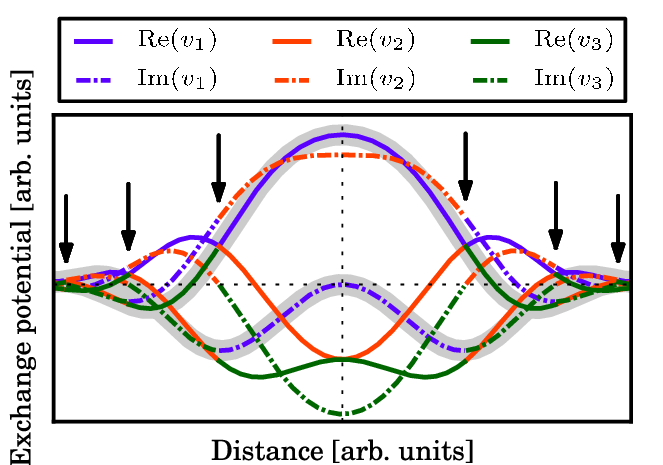}
  \caption{Real and imaginary parts of the 
  three branches $v_1$, $v_2$ and $v_3$ of the
    LDA exchange potential
    generated by a
    complex-scaled Gaussian density.
    The branches are stitched together 
    to form one continuous exchange potential,
    indicated by the shaded bands.
    Arrows indicate the branch points.}
  \label{fig:stitching}
\noindent{\color{JPCCBlue}{\rule{\columnwidth}{1pt}}}
\end{figure}

XC functionals $E_\XC^\theta[n^\theta]$ can be
derived by analytic continuation consistently with
Eqns.~\eqref{eq:csstates}--\eqref{eq:csoperators}.
Consider for example spin-paired LDA.
The XC energy is complex-scaled by rotating the integration contour
from the real axis into the complex plane:
\begin{align}
  E_\XC[n] = \int n(\ve r) \epsilon(n(\ve r)) \idee \ve r
  = \int n(\ve r\ee^{i\theta}) \epsilon(n(\ve r\ee^{i\theta}))
  \idee \ve r\, \ee^{i 3 \theta},
\end{align}
which is also a functional $E_\XC^\theta$ of $n^\theta(\ve r)$ by
\eqref{eq:csdensity}.
The potential follows
as $v_\XC^\theta(\ve r) = \delta E_\XC^\theta[n^\theta]/\delta n^\theta(\ve r)$.
Thus the exchange
part of the potential becomes
\begin{align}
  v_\X^\theta(\ve r) 
  = -\left(\frac 3 \pi \right)^{1/3} \ee^{-i\theta}[n^\theta(\ve r)]^{1/3}
  = v_\X(\ve r \ee^{i\theta}),
\end{align}
as in \eqref{eq:csoperators}.
Due to the complex cube root, the exchange potential is three-valued.
However the potential is the complex continuation of a corresponding
real potential for $\theta=0$, and so must remain continuous.
Further, since the complex scaling operation leaves the
origin $\ve r = 0$ unaffected, $v_\X^\theta(0)$ must be real and independent
of $\theta$.
Starting at $\ve r = 0$ we therefore evaluate $v_\X^\theta(\ve r)$ as
the principal branch of the cube root of the density.  For some $\ve
r$ the density may approach a branch point so the potential
becomes discontinuous.  This situation is illustrated
in Figure 
\ref{fig:stitching}, where the three branches,
evaluated from a Gaussian density 
with $\theta=0.5$, have different colors.
At a branch point, one can always choose another branch such that the
resulting, \emph{stitched} potential becomes continuous and yields the
correct energy $E_\X^\theta[n^\theta]$ which does not depend on $\theta$.

Consider next the Perdew--Wang
parametrization of
correlation\cite{PhysRevB.45.13244}.
The correlation potential is expressed in terms of the
Wigner--Seitz radius $\sqrt{r_s(\ve r)}\sim [n(\ve r)]^{-1/6}$ and takes the
form
\begin{align}
  v_c(r_s) = \epsilon_c(r_s) - \frac13 \diff{\epsilon_c(r_s)}{r_s} r_s,
\end{align}
where
\begin{align}
\epsilon_c(r_s) &= -2 A (1 + \alpha_1 r_s) \ln\left(1 + \frac{1}{Q_1(r_s)}\right),\\
  Q_1(r_s) &= 2 A \sum_{i=1}^4 \beta_i r_s^{i/2}.
\end{align}
Here $A$, $\alpha_1$ and $\beta_{1...4}$ are real constants.
This expression is straightforward to complex-scale using the
stitching method already presented.  We first evaluate
$\sqrt{r_s}$ at each point on the real-space grid by stitching
$[n^\theta(\ve r)]^{-1/6}$.  Then the complex
logarithm $\ln(1 + 1 / Q_1(r_s))$ is stitched to
obtain $\epsilon_c(r_s)$.
Other local or semilocal (GGA)
functionals can be similarly complex-scaled.

\begin{figure}
\noindent{\color{JPCCBlue}{\rule{\columnwidth}{1pt}}}
  \includegraphics{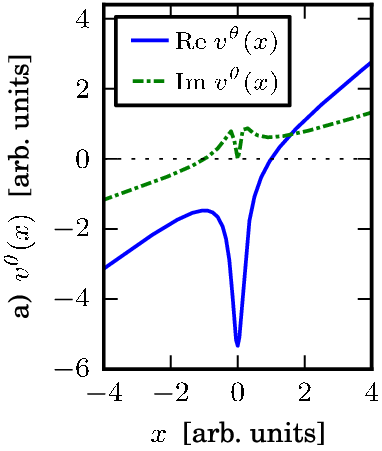}
  \includegraphics{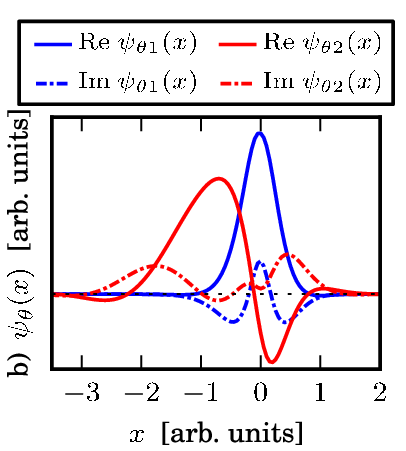}\\
  \includegraphics{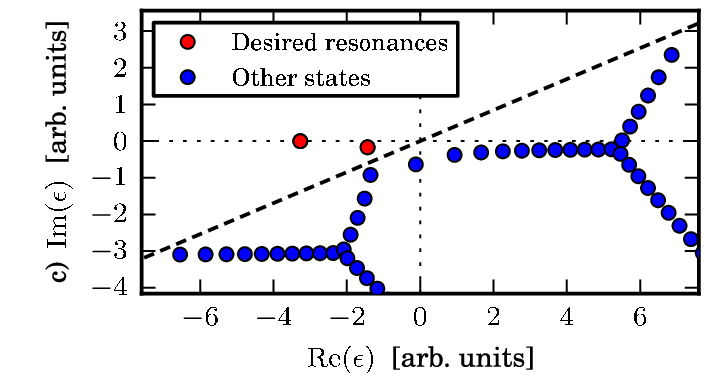}
  \caption{a) Real and imaginary parts of the complex-scaled
    potential.  b) Real and imaginary parts of the lowest
    resonance wavefunctions.  c) The spectrum.
    The dashed line $\arg z = \theta$ separates the two 
    real resonances (red) with imaginary parts of $-2.6\times10^{-3}$
    and $-0.17$ from
    the remaining eigenstates (blue) which are artifacts of the simulation box
    or continuum.}
  \label{fig:1d-example}
\noindent{\color{JPCCBlue}{\rule{\columnwidth}{1pt}}}
\end{figure}

One of the challenges in Kohn--Sham DFRT is to reliably determine which
states should be occupied, as the complex eigenvalues have no natural
ordering.  Consider independent particles in 1D near
an atom represented by a soft Coulomb potential with charge $Z$ in a
uniform electric field of strength $F$.  This system has the external
potential $F x\ee^{i\theta} - Z / \sqrt{x^2\ee^{i2\theta} + \alpha^2}$.
Figure
\ref{fig:1d-example} shows a) the complex-scaled potential, b) wavefunctions
and c) eigenvalues for $F=4$, $Z=4$, $\alpha=0.15$, and $\theta=0.4$ in a
simulation box of size $10$.  In the spectrum
on Figure 
\ref{fig:1d-example}(c), the dashed line $\arg z = \theta$ divides
the complex plane in two parts.
On the upper left side there are two
eigenvalues that correspond to physical resonances.  These states would
have been
bound if no electric field had been applied to the system, but are 
now situated just below the real axis.
Below the dashed line $\arg z = \theta$,
the spectrum forms a system of lines.
It has been demonstrated by Cerjan and
co-workers\cite{0022-3700-11-7-002,QUA:QUA560140408}
that the numerical range
(the set of values $(\psi^\theta | H_{\mathrm{stark}}^\theta | \psi^\theta)$
for all normalized states $\psi^\theta$) of the
complex-scaled Stark Hamiltonian, and thus its entire
continuous spectrum, falls within this region.
The discrete eigenvalues above the line $\arg z = \theta$ can therefore
be identified as originating from bound
states in the isolated atom, and
can now be assigned occupations
in order of increasing real (or negative imaginary) part of the
energy, while the remaining states are left unoccupied.
In practical calculations using iterative eigensolvers, particularly
when far away from self-consistency, eigenvalues originating from the
continuum may appear above the line $\arg z = \theta$.
As these states should not be occupied, we use a simple rule to identify them.
They are occupied in ascending
order of $\real \epsilon + \alpha (\imag \epsilon)^2$,
where $\alpha$ is a tunable parameter.
The value $\alpha=2/\sin \theta$ generally works for the atoms considered here.

\begin{figure}
\noindent{\color{JPCCBlue}{\rule{\columnwidth}{1pt}}}
  \includegraphics{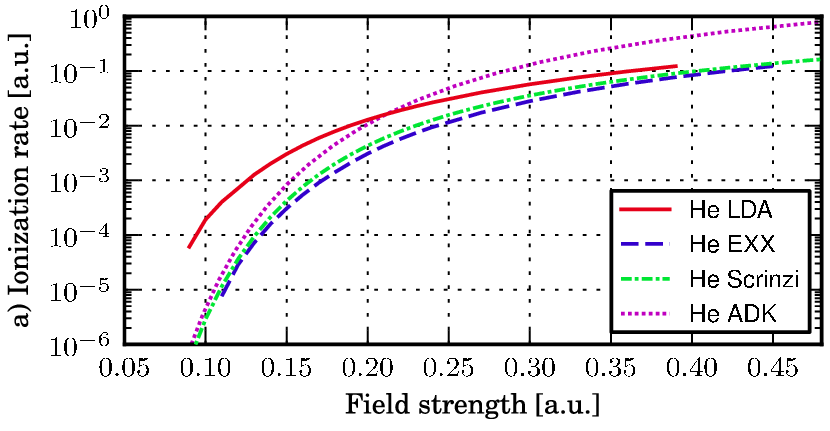}
  \includegraphics{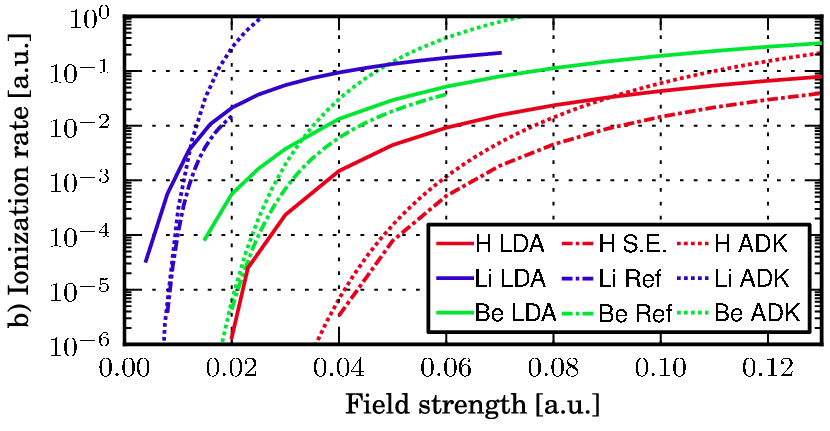}
  \caption{Ionization rates 
    ($1\,\textrm{a.u.} \approx 4.13 \times 10^{16}\,\mathrm{s}^{-1}$)
    of (a) He and (b) H, Li, and Be
  as a function of electric
  field strength ($1\,\textrm{a.u.} \approx 5.14\times 10^{11}\,\mathrm{V/m}$).
  Rates are calculated using DFRT (LDA or EXX) and the
  Ammosov--Delone--Krainov (ADK) method\cite{adk}.
  Rates of H are also calculated by solving the complex-scaled 
  Schrödinger equation.  Accurate reference rates 
  from first-principles methods are shown
  for He\cite{PhysRevLett.83.706},
  Li,\cite{PhysRevA.47.3122} and Be\cite{0953-4075-33-24-308}.
  }
  \label{fig:atom-ionization}
\noindent{\color{JPCCBlue}{\rule{\columnwidth}{1pt}}}
\end{figure}

Like in standard DFT calculations, it is the outermost (valence)
electrons that determine most properties of a system.
Nuclear point charges cause numerical
difficulties due to their central singularity.  Pseudopotentials
solve this problem by replacing the point charges by smooth
charge distributions, while making sure to account properly 
for the core--valence interaction.
Here we use the normconserving
Hartwigsen--Goedecker--Hutter (HGH) pseudopotentials\cite{PhysRevB.58.3641}.
They can be explicitly complex-scaled since they are parametrized
as polynomials and Gaussians.  In the calculations
below, we use the potentials that include all electrons as valence
electrons while only smoothening the nuclear potential
(Li and Be have 3 and 4 valence electrons, respectively).
With this choice the approach is demonstrated on systems with more than
one occupied Kohn--Sham state.  The approach is also compatible with
standard frozen-core pseudopotentials.

Figure 
\ref{fig:atom-ionization}(a) shows
the ionization rate for He as a function of electric field
strength calculated using various methods.  
The reference results
are based on direct solution of the
complex-scaled two-particle Schrödinger equation and thus represent
the closest to an exact calculation\cite{PhysRevLett.83.706}.
The DFRT rates $\Gamma$ for LDA and EXX
are obtained directly from \eqref{eq:complexenergy} after solving the
KS equations \eqref{eq:kohnsham} self-consistently for the complex
density using
$\theta=0.35$.  A value of $\theta$ is suitable if it is large enough
to localize the resonant KS states, and if the results converge rapidly
with grid spacing.
A very fine grid spacing of
0.08 a.u.\ is still needed to converge the lowest rates.  The LDA
substantially overestimates the ionization rate, particularly for small
fields, while EXX is in very good
agreement with the reference.  EXX results are obtained by setting the exchange
energy to minus half the Hartree energy, which is exact for two-electron
systems.

Also shown are results from the
Ammosov--Delone--Krainov (ADK) method\cite{adk}.  This is a
simple approximation for ionization rates in atoms, based on the atomic
ionization potential.
ADK is accurate for low fields because the ionization rate is strongly
linked to the ionization potential in this limit.  However it greatly
overestimates rates for large fields.
We attribute the inaccuracy of LDA for low fields to
its well-known underestimation of ionization potentials, taken as
minus the energy of the highest occupied Kohn--Sham
orbital
(0.57 Hartree from LDA, versus 0.92 from EXX and 0.90 from experiment).
This error of LDA is ultimately linked to the 
exponential rather than Coulomb-like decay of the
potential.\cite{PhysRevA.49.2421}
Note that more accurate ionization
potentials can be calculated by subtracting the total energy of the
charged and the neutral system.
Ionization rates based on this method have been presented
with CODFT\cite{doi:10.1021/jz3006805}.

Similarly calculated ionization rates with LDA and ADK are shown for
H, Li, and Be in Figure
\ref{fig:atom-ionization}(b) along with
reference values for H from ordinary one-particle calculations,
and for Li\cite{PhysRevA.47.3122} and Be\cite{0953-4075-33-24-308}.
Generally the atoms with lower atomization potentials have higher ionization
rates, and again a large discrepancy shows between ADK and LDA for low fields.
\begin{figure}
\noindent{\color{JPCCBlue}{\rule{\columnwidth}{1pt}}}
  \includegraphics{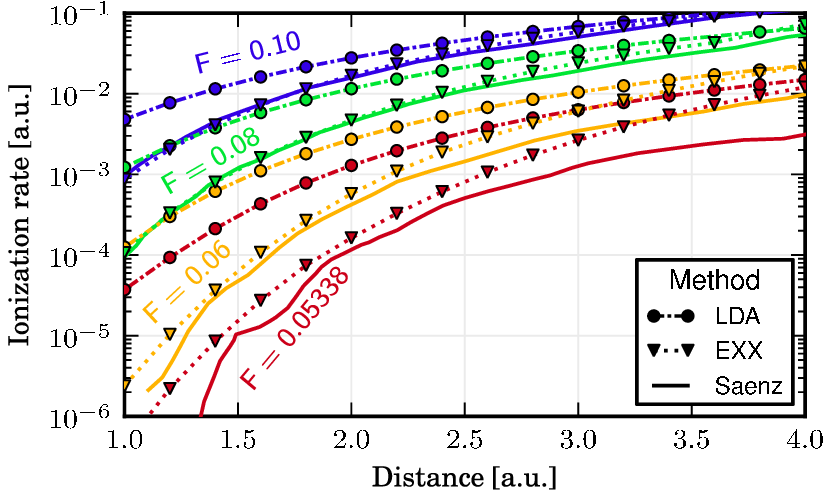}
  \caption{Ionization rate of H$_2$ as a function of internuclear distance.
    Different methods are denoted by different line and point styles, while colors denote different static field strengths in atomic units.
  }
  \label{fig:h2-dissociation}
\noindent{\color{JPCCBlue}{\rule{\columnwidth}{1pt}}}
\end{figure}

Figure
\ref{fig:h2-dissociation} shows ionization rates for the
H$_2$ molecule as a function of internuclear distance calculated for different
field strengths with LDA and EXX.  The molecular axis is
parallel to the electric field.
The nuclei are described as fixed point particles, so only the static
electron ionization yield is calculated.
The reference calculations by Saenz\cite{PhysRevA.61.051402} 
correspond to an accurate
solution of the two-particle complex-scaled Schrödinger equation.  

H$_2$
in the dissociation limit is a pathological case in DFT as the system
is dominated by strong static correlations that most functionals fail to
capture.
In this limit the system consists of two isolated, charge neutral
atoms.  A static calculation with an electric field
will produce
a different solution where both electrons reside on the atom favored by the
field, although the situation at intermediate distances as here is more
complicated.

LDA again overestimates ionization rates, particularly for short bond
lengths.  For large field strength and short bond lengths, the
agreement between EXX and the reference is almost perfect.
However ionization rates on the order of $10^{-5}$ a.u.\ lose accuracy due to
the numerical dependence of energy on
$\theta$.  This error can be eliminated by optimizing the choice of $\theta$
and using a more fine grid spacing\cite{QUA:QUA560140408}
(0.1 Bohr with $\theta=0.22$ for the data points in question).
We attribute most of the
disagreement at short bond lengths between EXX and the reference to
this error.

At large bond lengths and large field strength, the EXX agrees well
with the reference.  This corresponds to the case where both electrons
reside mostly on the same atom.  For smaller field strengths
the system corresponds more closely to the strongly correlated case,
and the error is larger.

The accuracy of the XC approximation is clearly a determining factor for
the quantitative success of
DFRT.  We have here considered very simple functionals, and
in particular LDA exhibits large errors.  Phenomena of excited states
depend intricately on the decay properties of the potential far from the
system, which are difficult to describe with semilocal functionals.
A promising method to solve this problem is to introduce
a fictitious ``XC density''
which defines a correction to the XC potential, giving it
a Coulomb-like decay.\cite{PhysRevLett.107.183002}
This can greatly improve the
accuracy of ionization rates.
We expect the derivation of improved XC functionals for
DFRT to be one of the next major
steps in the development of this method.

The presented calculations demonstrate
the reliability and performance of DRFT
for realistic atoms and dimers.  The extension to other molecular
systems and nanostructures is straightforward, opening the path
towards a systematic study of the electronic and structural properties
of metastable complexes.  Furthermore, DFRT has
implications for the discussion and analysis of resonances in
molecular electronics as well as to the description of intermediates in
surface--molecule interactions, as the method introduces decay
processes in a natural way into the widely used first-principles
density-functional framework.

A future goal is to enable DFRT calculations for time-dependent systems,
where time propagation can be started from statically determined
resonant states.
This paves the road to tackle dynamical processes through metastable
intermediates, as seen in the recently available
ultrafast and ultraintense laser probes that allow to extract temporal
and spatial information of electron and ion
dynamics\cite{tunnellingtime} as well as
imaging.\cite{zhouimaging,PhysRevLett.109.133202}
\titleformat{\section}{\bfseries\sffamily\color{JPCCBlue}}{\thesection.~}{0pt}{\large$\blacksquare$\normalsize~}

\section*{AUTHOR INFORMATION}
\subsubsection*{Corresponding Authors}
\noindent *E-mail: asklarsen@gmail.com (A.H.L.);\\umberto.degiovannini@ehu.es (U.D.G.);\\awasser@purdue.edu (A.W.); \\angel.rubio@ehu.es (A.R.)
\subsubsection*{Notes} 
\noindent The authors declare no competing financial interest.

\section*{ACKNOWLEDGMENTS} 
We acknowledge funding from:
The European Research Council Advanced Grant DYNamo (ERC-2010-AdG
Proposal No.\ 267374),
Grupo Consolidado UPV/EHU del Gobierno Vasco (IT578-13),
the European Commission (Grant number 280879-2
CRONOS CP-FP7),
and the Spanish grants FIS2010-21282-C02-01 and PIB2010US-00652.
DLW and AW acknowledge funding from the U.S.\ National Science
Foundation CAREER program Grant No.\ CHE-1149968.

\end{document}